# Modeling of dielectric hysteresis loops in ferroelectric semiconductors with charged defects.

Anna N. Morozovska[1], Eugene A. Eliseev[2]

[1]V.Lashkaryov Institute of Semiconductor Physics, NAS of Ukraine,
45, pr. Nauki, 03028 Kiev, Ukraine,
e-mail: morozo@i.com.ua

[2]Institute for Problems of Materials Science, National Academy of Science of Ukraine,
Krjijanovskogo 3, 03142 Kiev, Ukraine,
e-mail: eliseev@i.com.ua

We have proposed the phenomenological description of dielectric hysteresis loops in ferroelectric semiconductors with charged defects and prevailing extrinsic conductivity. Exactly we have modified Landau-Ginsburg approach and shown that the macroscopic state of the aforementioned inhomogeneous system can be described by three coupled equations for three order parameters. Both the experimentally observed coercive field values well below the thermodynamic one and the various hysteresis loop deformations (constricted and double loops) have been obtained in the framework of our model. The obtained results quantitatively explain the ferroelectric switching in such ferroelectric materials as thick PZT films.

## 1. Introduction

The spontaneous electric displacement switching under the external field is one of the essential features of the ferroelectric materials [1], [2]. The main comprehensively studied characteristics of ferroelectric hysteresis loop are the spontaneous displacement and coercive field values [2], [3]. However, conventional theoretical approaches give significantly incomplete picture of the displacement switching, namely:

- Nucleation theory [4] proved that the ferroelectrics inhomogeneities promote the domains nucleation, appearance of several nuclei in the ferroelectric film do not need to overcome the energetic barrier due to the long range interaction between them [5]. Calculated value of the coercive field is inversely proportional to the film thickness [6] and thus nucleation theory can explain the drastic increase of coercive field with film thickness decrease experimentally observed in some thin films [7], [8]. However, nucleation theory predicts that coercive field tends to zero at film thickness increasing, and this theory seems unsuitable for thick films and bulk materials.

- The domain kinetics theory evolved in [9], [10] uses Kolmogorov-Avrami statistical model (see chapter 4 in [11]). It allows one to describe experimental data with high accuracy due to the great amount of fitting parameters without clear interpretation. The obtained results represent themselves only computer simulations.

- Landau-Ginsburg theory (see [3], chapter 7 in [11]) evolved for the mono-domain perfect ferroelectrics describes switching without domains (so called homogeneous switching) and neither domain pinning, nor domain nucleation and domain movement. Calculated values of thermodynamic coercive field [2], [3] are from several times to several orders greater than the experimentally observed values for the real bulk ferroelectric materials (see e.g. [12], [13]).

Renormalized Landau-Ginsburg free energy with coefficients depending on the film thickness and boundary conditions characteristics [14], [15] does not describe adequately ferroelectric hysteresis even in epitaxial film, because it predicts very sharp coercive field relaxation up to the aforementioned thermodynamic coercive field with film thickness increase [16].

• As for the imperfect doped ferroelectrics, here the hysteresis loops look much thinner [17]-[20] than Landau-Ginsburg ones, and its shape undergoes various deformations ("slim" loops [21], chapter 5 in [12]; minor loops [22], [23]; constricted loops [24], [25]). Therefore, the modification of Landau-Ginsburg free energy (and thus Landau-Khalatnikov equation [3] derived from it by variational method) suitable for the description of inhomogeneous ferroelectrics switching seems necessary. Incorporation of the modified free energy with inhomogeneities contribution could be a solution to this problem.

To our mind, the inhomogeneous electric fields caused by charged defects should be taken into account. However in contrast to the random field theory, developed for the relaxor insulator ferroelectrics [26], semiconductor properties of these materials [27] (at least extrinsic conductivity created by charged impurities [28], [29]) should be taken into consideration. Right in the way we modified Landau-Ginsburg approach for the ferroelectrics-semiconductors with charged defects and found both the essential coercive field decrease and the hysteresis loops deformation experimentally observed in inhomogeneous systems.

## 2. The problem

The majority of imperfect ferroelectrics with non-isovalent impurities or some unavoidable imperfections would be considered rather as extrinsic semiconductors [27], [28] than perfect insulators.

We assume that almost immovable non-stechiometric defects or non-isovalent impurity centers are embedded into hypothetical perfect uniaxial ferroelectric (z direction coincides with the polar axis). We suppose that impurity centers or defects are ionized (e.g. after UV, photo- or thermal excitation) and even in the absence of proper conductivity, they provide a prevailing extrinsic conductivity in the bulk sample.

For the sake of simplicity, we regard that the sample as a whole is the electro-neutral *n*-type extrinsic semiconductor with positively charged defects with density $\rho_s(\mathbf{r})$. The microscopic spatial distribution of these defects charge density $\rho_s(\mathbf{r})$ is characterized by the average charge density $\overline{\rho}_s$ proportional to the ionized defects concentration and microscopic modulation $\delta\rho_s$, i.e. $\rho_s(\mathbf{r}) = \overline{\rho}_s + \delta\rho_s(\mathbf{r})$. The modulation $\delta\rho_s$ fluctuates due to the great variety of misfit effects (lattice distortions, local shift, possible clusterization at high defect concentration). For the sake of simplicity, let us assume that charged defects spatial distribution is quasi-homogeneous, i.e. the average distance between defects $d \cong a/\sqrt[3]{n_d}$ (a is lattice constant, $n_d$ is defect concentration). The size distribution function of charged defects is well localized near its sharp maximum at the average size $2r_0$.

The movable screening clouds $\delta n(\mathbf{r},t)$ surround each charged center (see Scheme 1a). The characteristic size of these screening clouds is of the same order as the Debye screening radius $R_D$. When one applies the external field $E_0$, screening clouds of free charges are deformed, and nano-system "defect center + screening cloud" becomes polarized (Scheme 1b). Polarized regions "$\delta\rho_s(\mathbf{r})+\delta n(\mathbf{r},t)$" cause the additional inner electric field fluctuations $\delta E(\mathbf{r},t)$. According to the equations of state, the fluctuations of the inner electric field $\delta E$ cause displacement fluctuations $\delta D$.

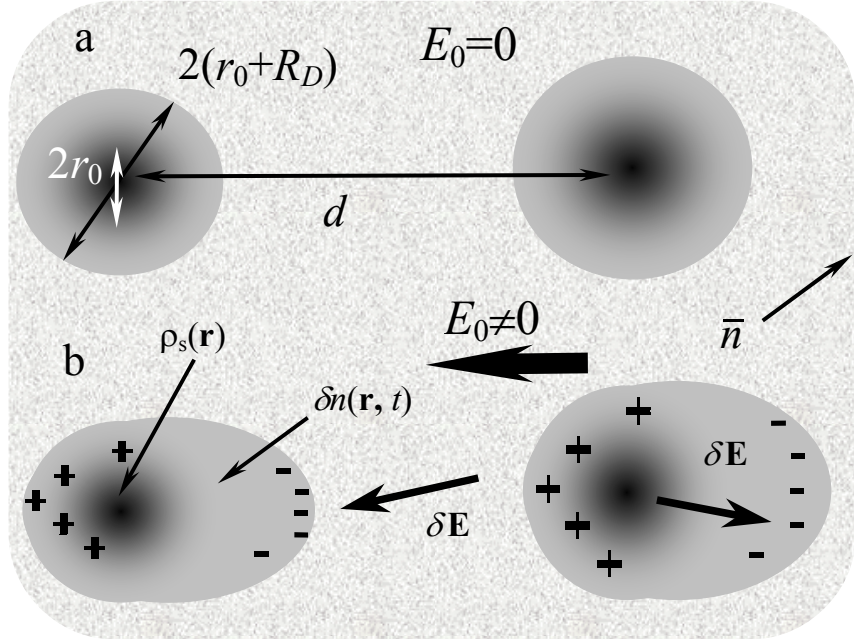

**Scheme 1.** The charged defects with the charge density $\rho_s$ (dark circles with radius $r_0$) are screened by the free charges with density $\delta n$ (gray circles or ellipses with screening radius $R_D$). The parts "a" and "b" show the system with the zero and nonzero external field $E_0$ respectively.

### 3. General equations

Hereinafter we assume that the period of external field changing is much greater than the free carriers relaxation time, and therefore the quasi-static approximation $rot\,\mathbf{E}=0, \quad rot\,\mathbf{H}=0$ is valid. Thus, Maxwell's equations for the quasi-static electric field $\mathbf{E}$ and displacement $\mathbf{D}$ have the form:

$$div\,\mathbf{D}=4\pi\rho, \quad \frac{\partial \mathbf{D}}{\partial t}+4\pi\,\mathbf{j}_c=0 \quad . \tag{1}$$

They have to be supplemented by the material expressions for current and charge density:

$$\mathbf{j}_c=\sum_m\left(\mu_m\rho_m\mathbf{E}-\kappa_m\,grad\,\rho_m\right), \quad \rho(\mathbf{r},t)=\sum_m\rho_m(\mathbf{r},t)+\rho_s(\mathbf{r}). \tag{2}$$

Here $\rho_m$, $\mu_m$ and $\kappa_m$ are the $m$-type movable charge volume density ($m=n,p$), mobility and diffusion coefficient respectively, $\mathbf{j}_c$ is the macroscopic free-carriers current, $\rho_s(\mathbf{r})$ is the fluctuating charge density of static defects.

Keeping in mind that the sample is the extrinsic semiconductor with prevailing $n$-type conductivity, hereinafter we neglect the proper conductivity, put $n\approx\sum_m\rho_m$ and omit the subscript "$m$". So equations (1), (2) can be rewritten as:

$$div\mathbf{D}=4\pi(n+\rho_s), \quad \mu\,n\,\mathbf{E}-\kappa\,grad\,n+\frac{1}{4\pi}\frac{\partial}{\partial t}\mathbf{D}=0. \tag{3}$$

Here $n<0$, $\mu<0$ and $\kappa>0$. In accordance with Einstein relation $\mu/\kappa\approx e/k_BT^*$ ($e<0$) the Debye screening radius $R_D=\sqrt{\kappa/4\pi n\mu}\approx\sqrt{\left|k_BT^*/4\pi ne\right|}$ [27], [28]. Hereafter we suppose that homogeneous external field $E_0(t)$ is applied along polar $z$-axis. The infinite in the transverse directions rather thick sample with equivalent boundaries occupies the region $-\ell<z<\ell$. We suppose that the potentials of electrodes are given, so that the inner field satisfies the condition:

$$\frac{1}{2\ell}\int_{-\ell}^{\ell} E_z(\mathbf{r},t)\,dz = E_0(t)\,. \tag{4}$$

Hereinafter we introduce the averaging over sample volume $V$:

$$f(\mathbf{r},t) = \overline{f(t)} + \delta f(\mathbf{r},t), \quad \overline{f}(t) = \frac{1}{V}\int_{V} f(\mathbf{r},t)d\mathbf{r}\,. \tag{5}$$

Hereinafter the dash designates the averaging of functions $f = \{n,\,\rho_s,\,E,\,D\}$, by definition $\overline{\delta f(\mathbf{r},t)} = 0$. The spatial distribution of the deviations $\delta f(\mathbf{r},t)$ consist of the part caused by spontaneous displacement screening [27] and localized in the ultrathin screening surface regions [30] and the one caused by microscopic modulation $\delta\rho_s$ with quasi-homogeneous spatial distribution. For the $\mu$m-thick sample the contribution from the ultrathin screening region to the average functions is negligibly small, and the averaging over sample volume is equivalent to the statistical averaging with homogeneous distribution function. Using this distribution function properties, one can regard that:

$$\overline{\delta f^{2m+1}(\mathbf{r},t)} \approx 0, \qquad m = 1;2\dots\,, \tag{6}$$

and the correlation between the different $\delta f$-functions is equal to zero if the total power of the functions is an odd number.

It follows from (3)-(5) that:

$$\mathbf{D}(\mathbf{r},t) = \mathbf{e}_z\,\overline{D(t)} + \delta\mathbf{D}(\mathbf{r},t)\,, \tag{7a}$$

$$\mathbf{E}(\mathbf{r},t) = \mathbf{e}_z E_0(t) + \delta\mathbf{E}(\mathbf{r},t)\,. \tag{7b}$$

Here $\mathbf{e}_z$ s the unit vector directed along z-axis, $\overline{E}$ is the applied uniform field $E_0(t)$ and $\overline{E}_{x,y} = 0$. Notice that the average values $\overline{E}$, $\overline{D}$ are determined experimentally [1]-[3] most of the times. Having substituted (7) into (3) and averaged, one can obtain the expressions for the average quantities, namely:

$$\overline{n} = -\overline{\rho}_s, \qquad \mu\,\overline{n}\,E_0 + \mu\overline{\delta n\,\delta E_z} + \frac{\partial}{\partial t}\frac{\overline{D}(t)}{4\pi} = 0, \qquad \mu\overline{\delta n\,\delta E_{x,y}} = 0\quad. \tag{8}$$

The absence of the space charge average density $\overline{\rho}$ follows from the sample electro-neutrality. Using $\rho_s = \overline{\rho}_s + \delta\rho_s$, $n = \overline{n} + \delta n$, (7) and (8) one can obtain from (3) that

$$div(\delta\mathbf{D}) = 4\pi(\delta n + \delta\rho_s(\mathbf{r}))\,, \tag{9}$$

$$\mu\,\left(\delta n\,E_0\mathbf{e}_z - \overline{\rho}_s\delta\mathbf{E}\right) + \mu\!\left(\delta n\delta\mathbf{E} - \overline{\delta n\,\delta\mathbf{E}}\right) - \kappa\,\,grad\,\delta n + \frac{1}{4\pi}\frac{\partial}{\partial t}\delta\mathbf{D} = 0\,. \tag{10}$$

The term $\mu\!\left(\delta n\,\delta\mathbf{E} - \overline{\delta n\,\delta\mathbf{E}}\right)$ in (10) can be interpreted as the fluctuating circular electric currents around charged defects, which do not contribute into the average macroscopic current.

As the equation of state, we use the Landau-Khalatnikov equation for the displacement z-component relaxation, but take into account the influence of fluctuating electric field $\delta E_z$ created by charged defects and correlation effects. This approach takes into consideration the spatial-temporal dispersion of the ferroelectric material. Using the original approach evolved in paper [31] and formula (7), we modify classical Landau-Khalatnikov equation $\Gamma\dfrac{\partial}{\partial t}D_z + \alpha\,D_z + \beta D_z^3 = E_z$ and obtain the following system of coupled equations:

$$\Gamma\frac{\partial\overline{D}}{\partial t} + \left(\alpha + 3\beta\overline{\delta D^2}\right)\overline{D} + \beta\overline{D}^3 = E_0(t)\,, \tag{11}$$

$$\Gamma\frac{\partial}{\partial t}\delta D + \left(\alpha + 3\beta\overline{D}^2\right)\delta D + 3\beta\overline{D}\left(\delta D^2 - \overline{\delta D^2}\right) + \beta\delta D^3 - \gamma\frac{\partial^2\delta D}{\partial r^2} = \delta E_z. \quad (12)$$

Here $\Gamma>0$ is the kinetic coefficient, $\alpha<0$, $\beta>0$, $\gamma>0$ are parameters of the hypothetical pure (free of defects) sample. Hereafter we denote $\delta D_z \equiv \delta D$. We would like to underline, that the sum of equations (11)-(12) coincides with Landau-Khalatnikov equation [3] for $D$ only under the condition $\delta\rho_s$=0. This condition corresponds to the absence of inhomogeneities or their seeding, and thus only the homogeneous polarization switching can take place when the external field exceeds thermodynamic coercive field. In such case the inner field in the bulk of the sample is the sum of the external field and predetermined depolarization field originated from displacement screening [27]. In contrast to this at $\delta\rho_s\neq0$ the inner field $\delta\mathbf{E}$ contains the random component dependent over $\delta\rho_s(\mathbf{r})$ and $\delta D$ in accordance with (9)-(10).

The system of equations (9)-(12) is complete, because the quantities $\delta n, \delta\mathbf{E}$ can be expressed via the fluctuations of displacement $\delta D$ and $\delta\rho_s(\mathbf{r})$ allowing for (9), (10). The spatial distribution and temporal evolution of the displacement $\mathbf{D}(\mathbf{r},t)$ in the bulk sample is determined by the non-linear system (9)-(12) supplemented by the initial distributions of all fluctuating variables (e.g. distribution of charged defects). However, in our opinion, only the adopted for the stochastic differential equations implicit numerical schemes can be used in order to obtain the numerical solutions of (9)-(12). Therefore, hereinafter we consider only the average characteristics of the inhomogeneous ferroelectric semiconductors. The study of the mechanisms of domain wall pinning by the given distribution of charged defects, domain nucleation during spontaneous displacement reversal is beyond the scope of this paper. Similar problems for the ferroelectrics - ideal insulators were considered in details earlier (see e.g. [32], [33]).

## 4. Coupled equations.

In order to simplify the nonlinear system (9)-(12) the following assumption has been used. The charged inhomogeneities are surrounded by screening clouds (see Fig. 1), therefore $\delta\rho_s \sim \delta n$ and

$$\overline{\delta\rho_s\delta n} \approx -\eta\overline{\delta\rho_s^2}, \qquad\qquad 0 < \eta << 1. \quad (13)$$

Small positive function $\eta$ is determined by the ratio $R_D/r_0$ (see Fig. 1 and Appendix A). When external field amplitude increases, $\eta$ value slightly decreases due to the polarization of the system "charged fluctuation + screening cloud". We suppose that $\eta$ can be approximated by effective constant value at not very high external field.

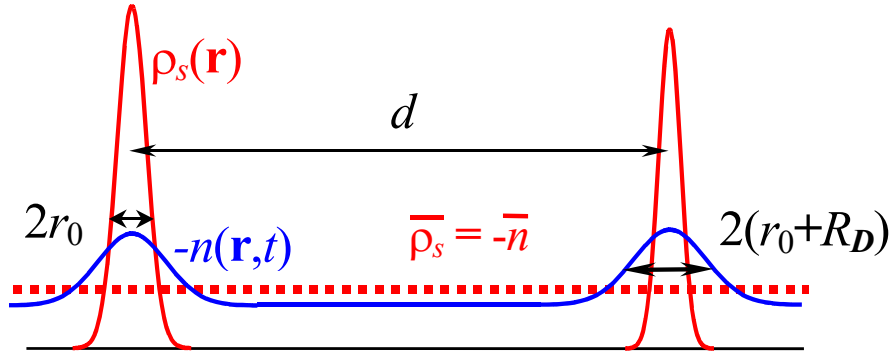

**Figure 1.** The screening of the charged defects $\delta\rho_s$ by free charges $\delta n$ at small amplitude of external field.

After elementary transformations of (10), the equation for electric field fluctuation $\delta \mathbf{E}$ acquires the form:

$$\delta \mathbf{E} = \frac{1}{4\pi\mu\bar{\rho}_s} \frac{\partial}{\partial t} \delta \mathbf{D} + \mathbf{e}_z \frac{\delta n}{\bar{\rho}_s} E_0(t) - \frac{\kappa}{\mu\bar{\rho}_s} \, grad \, \delta n + \frac{\delta n \, \delta \mathbf{E} - \overline{\delta n \, \delta \mathbf{E}}}{\bar{\rho}_s}. \qquad (14)$$

The equation (9) gives

$$\delta n = \frac{1}{4\pi} div(\delta \mathbf{D}) - \delta \rho_s(\mathbf{r}). \qquad (15)$$

Using (15) and (14) the electric field fluctuations $\delta \mathbf{E}$ caused by charged defects can be expressed via $\delta D$ and $\delta \rho_s$:

$$\delta \mathbf{E} \approx \frac{\partial}{\partial t} \frac{\delta \mathbf{D}}{4\pi\mu\bar{\rho}_s} + \left( \mathbf{e}_z E_0(t) - \frac{\kappa \, grad}{\mu} \right) \left( \frac{div(\delta \mathbf{D})}{4\pi\bar{\rho}_s} - \frac{\delta \rho_s}{\bar{\rho}_s} \right) + $$
$$+ \frac{\overline{\delta \mathbf{E} \, div(\delta \mathbf{D})} - \delta \mathbf{E} \, div(\delta \mathbf{D})}{4\pi\bar{\rho}_s} + \frac{\delta \mathbf{E} \delta \rho_s - \overline{\delta \mathbf{E} \delta \rho_s}}{\bar{\rho}_s} \qquad (16)$$

The equations (11), (12), and (16) is the self-consistent system of the nonlinear integral-differential equations for $\delta D$, $\delta E$ and $\bar{D}$. Its non-homogeneity is proportional to charge fluctuations $\delta \rho_s$ and external field $E_0$.

The approximate system of first-order differential equations for average displacement $\bar{D}$, its mean-square fluctuation $\overline{\delta D^2}$ and correlation $\overline{\delta D \delta \rho_s}$ can be derived after some elementary transformations of (11), (12) (see [34], Appendix B and (13), (16)). Thus, we obtain three coupled equations:

$$\Gamma \frac{\partial \bar{D}}{\partial t} + \left( \alpha + 3\beta \overline{\delta D^2} \right) \bar{D} + \beta \bar{D}^3 = E_0(t), \qquad (17a)$$

$$\frac{\Gamma_R}{2} \frac{\partial}{\partial t} \overline{\delta D^2} + \left( \alpha_R + 3\beta \bar{D}^2 \right) \overline{\delta D^2} + \beta \left( \overline{\delta D^2} \right)^2 = -E_0(t) \frac{\overline{(\delta \rho_s \delta D)}}{\bar{\rho}_s} + \vartheta \left( \overline{\delta \rho_s^2} \right), \qquad (17b)$$

$$\Gamma_R \frac{\partial}{\partial t} \overline{\delta D \delta \rho_s} + \left( \alpha_R + 3\beta \bar{D}^2 + \beta \overline{\delta D^2} \right) \overline{\delta D \delta \rho_s} = -E_0(t) \frac{\overline{\delta \rho_s^2}}{\bar{\rho}_s} \eta. \qquad (17c)$$

Here the renormalized coefficients $\alpha_R = \alpha + \left( \gamma + R_D^2 \right) / d^2$ and $\Gamma_R = \Gamma - 1/4\pi\mu\bar{\rho}_s$ have been introduced. The renormalization of $\alpha$ takes into account the finite value of correlation length and classical renormalization of gradient term $\gamma$ by Debye screening radius $R_D$ in the bulk of a sample [30]. The renormalization of $\Gamma$ takes into account the finite value of maxwellian relaxation time $\tau_m = -1/4\pi\mu\bar{\rho}_s$.

The additional source of displacement fluctuations is the term $\vartheta \left( \overline{\delta \rho_s^2} \right) = \left( 4\pi R_D \right)^2 \left( 1 - \eta \right) \overline{\delta \rho_s^2}$ in the right hand-side of (18b). It is originated from diffusion field $\kappa \, grad(n)/\mu\rho_s$ (see (10), (14) and definition $R_D = \sqrt{\kappa/4\pi\mu}$). This term has been neglected in the classical equations of Landau-Khalatnikov type as well as in our previous paper [35].

The system (17) determines the temporal evolution of the bulk sample dielectric response and have to be supplemented by the initial values of $\bar{D}$, $\overline{\delta D^2}$ and $\overline{\delta D \delta \rho_s}$ at $t$=0.

Coupled equations (17) have the following physical interpretation (compare with modified approach [31]). The macroscopic state of the bulk sample with charged defects can be described by three parameters: $\overline{D}$, $\overline{\delta D^2}$ and $\overline{\delta D \delta \rho_s}$. The long-range order parameter $\overline{D}$ describes the ferroelectric ordering in the system, and the disorder parameter $\overline{\delta D^2}$ describes disordering caused by inner electric fields caused by charged non-homogeneities $\delta \rho_s$. The correlation $\overline{\delta D \delta \rho_s}$ determines the correlations between the movable screening cloud $\delta n$ and static charged defects $\delta \rho_s$.

We would like to note that derived system of coupled equations (17) might possess chaotic regions, strange attractors as well as strongly non-ergodic behaviour and continuous relaxation time spectrum [11]. Any new system of such type demands a separate detailed mathematical study that was not the aim of this paper.

## 5. Ferroelectric hysteresis.

In this section, we consider the equilibrium solution of (17), which corresponds to the quasi-static external field changing. Let us demonstrate, how the ferroelectric hysteresis loop (i.e. the dependence of displacement $\overline{D}$ over the external field $E_0$) changes its shape under the presence of charged defects. First, let us rewrite equations (17) in dimensionless variables:

$$\frac{dD_m}{d\tau} - (1 - 3\Delta_D^2)D_m + D_m^3 = E_m \,,$$

$$\frac{\tau_R}{2}\frac{d\Delta_D^2}{d\tau} - (\xi - 3D_m^2)\Delta_D^2 + \Delta_D^4 = -E_m K_{D\rho} + g\,R^2 \,, \qquad (18)$$

$$\tau_R \frac{dK_{D\rho}}{d\tau} - (\xi - 3D_m^2 - \Delta_D^2)K_{D\rho} = -E_m R^2 \,.$$

Here, $D_m = \overline{D}/D_S$, $D_S = \sqrt{-\alpha/\beta}$, $E_m = E_0/(-\alpha D_S)$, $\Delta_D = \sqrt{\overline{\delta D^2}}/D_S$, $K_{D\rho} = \overline{\delta D \delta \rho_s}/\overline{\rho}_s D_S$, $R^2 = \eta\,\overline{\delta \rho_s^2}/\overline{\rho}_s^2$, $\tau = t/(-\alpha\Gamma)$, $\tau_R = \Gamma/\Gamma_R$ and $g = \dfrac{16\pi^2 R_D^2 \cdot \overline{\rho}_s^2}{-\alpha D_S^2}\dfrac{(1-\eta)}{\eta}$, $\xi = \alpha_R/\alpha$ at $\alpha < 0$.

The renormalized coefficient $\alpha_R = \alpha + \left(\gamma + R_D^2\right)\big/d^2$ is positive and $\alpha_R >> |\alpha|$ for the typical values of parameters. For example, in Pb(Zr,Ti)O3 in CGSE system $\gamma \approx 5 \cdot 10^{-16} cm^2$ [36], lattice constant $a \approx 4 \cdot 10^{-8} cm$, $\alpha \sim -(0.4 \div 2) \cdot 10^{-3}$ [13], 0.01-1% concentration of defects provides $n_d = 1.6 \times (10^{18} \div 10^{20}) cm^{-3}$ and so $\rho_s = 7.7 \times (10^8 \div 10^{10}) Q_{CGSE}/cm^3$, $d \cong a/\sqrt[3]{n_d} = (2 \div 9) \cdot 10^{-7} cm$, $R_D \sim \left(4 \div 8\right) \cdot 10^{-8} cm$ [28], defect radius $r_0 \overset{\sim}{<} a$ and so $\alpha_R \sim +(0.05 \div 20) \cdot 10^{-2}$. For the aforementioned values of parameters $\alpha$, $R_D$ and for hypothetical perfect material $D_S \sim (50 \div 100)\mu C/cm^2 \sim \left(15 \div 30\right) \cdot 10^4\,Q_{CGSE}/cm^2$, $\eta \sim \left(10^{-3} \div 10^{-2}\right)$ one obtains that $g \sim (3 \cdot 10^{-3} \div 2 \cdot 10^2)/\eta \sim (10^{-2} \div 10^4)$, $\xi \sim -(0.25 \div 400)$. At $d \sim (5 \div 10)r_0$ and $\eta \leq 0.01$ one obtains $R^2 = \eta(7 \div 60) \leq 1$. (see (A.7) in Appendix A). Below we numerically analyze the solution of the system (18) for the following parameters $0 \leq R \leq 1$, $0.5 < g < 50$, $-1 \leq \xi \leq -100$.

The equilibrium solution of (18) corresponds to the quasi-static external field changing. For the harmonic modulated applied field $E_m = E_{m0} \sin(w\tau)$ the dimensionless frequency $w = -\alpha\Gamma\omega$ must be much less than unity, we choose $w \sim (10^{-4} \div 10^{-2})$ and $\Gamma \approx \Gamma_R$. At such low frequencies, the initial conditions determine only the initial curve. They do not play any important role in the loop shape (hereinafter we put them zero: $D_m(\tau=0)=0$, $\Delta^2{}_D(\tau=0)=0$, $K_{D\rho}(\tau=0)=0$).

The typical hysteresis loops $D_m(E_m)$ obtained at $\xi \sim -1$, $R \sim 0.5$ and different small $g\sim1$ values are shown in Fig. 2. The increase of g value firstly leads to the coercive field decrease then to the constriction appearance, subsequent transformation to the double loop and finally to the loop disappearance. The broadening, slight tilting and smearing of the loop shape under applied field frequency increasing can be explained by dielectric losses increasing. It should be noted, that for $R>1$ and $\xi \to 1$, g=0, $w \to 0$ hysteresis loops are absent [34].

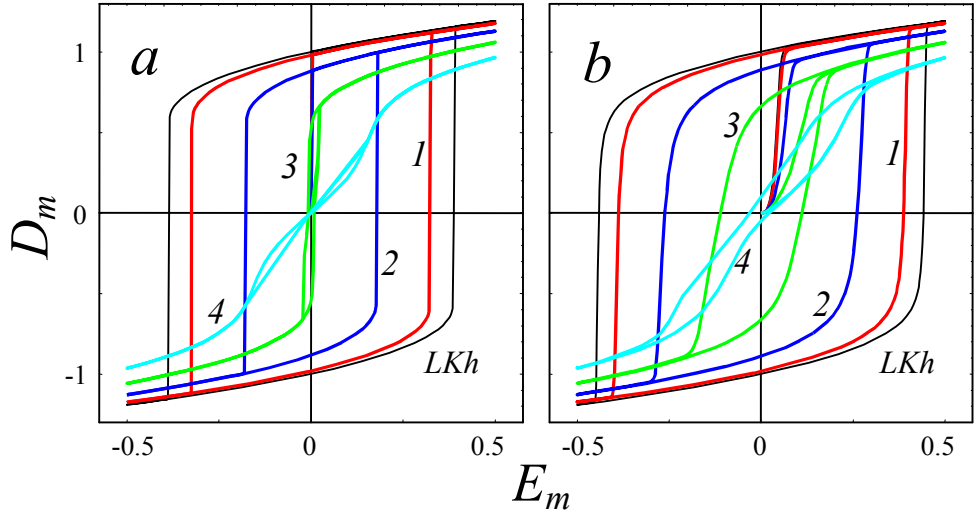

**Figure 2**. Hysteresis loops at frequencies $w$=2×10⁻⁴ (**plot a**), 2×10⁻² (**plot b**), $R^2 = 0.5$, $\xi$=-1 and for different values of $g$: 0.1 (curves 1), 0.5 (curves 2), 1 (curves 3), 1.5 (curves 4) and Landau-Khalatnikov loop (curve LKh).

Fig. 3 demonstrates the typical changes of hysteresis loop shape caused by increasing of charged defects density $\overline{\rho}_s$ (note that $g \sim \overline{\rho}_s$).

At high negative values $\xi \ll -1$ (see inset $a$), the increase of $g$ value firstly leads to the significant coercive field decrease, then to the loop disappearance. Namely, the coercive fields $E_{cm} \approx 0.0086$ and $E_{cm} \approx 0.0024$ for the loops 2, 3 in the basic plot are much smaller than its thermodynamic value $E_{cm} \approx 0.39$ for the LKh-loop in the inset $a$. The loop becomes much slimmer than lower under charged defects density $\overline{\rho}_s \sim g$ increasing. This effect is somewhat similar to the known "square to slim transition" of the hysteresis loops in some relaxor ferroelectrics [21]. Basic plot demonstrates the drastic decrease of coercive field for $R$=0.1, $\xi = -10$ and $g > 30$. At small values $|\xi| \leq 1$ (see inset $c$), the increase of g value firstly leads to the coercive field decrease then to the constriction appearance, subsequent transformation to the double loop and finally to the loop disappearance. Inset b demonstrates that disorder parameter $\Delta_D \neq 0$ over the region of hysteresis. Moreover, it reaches its maximum value near the coercive field, where $\overline{D} \to 0$. This means that coupled system (17) reveals **inhomogeneous** displacement switching, e.g. sample non-polarized as a whole splits into the oppositely polarized regions when external field reaches coercive value. In contrast to this, the Landau-Khalatnikov equation describes homogeneous switching with $\delta D \equiv 0$ independently on the external field. Therefore, system (17) is not equivalent to the equations of Landau-Khalatnikov type.

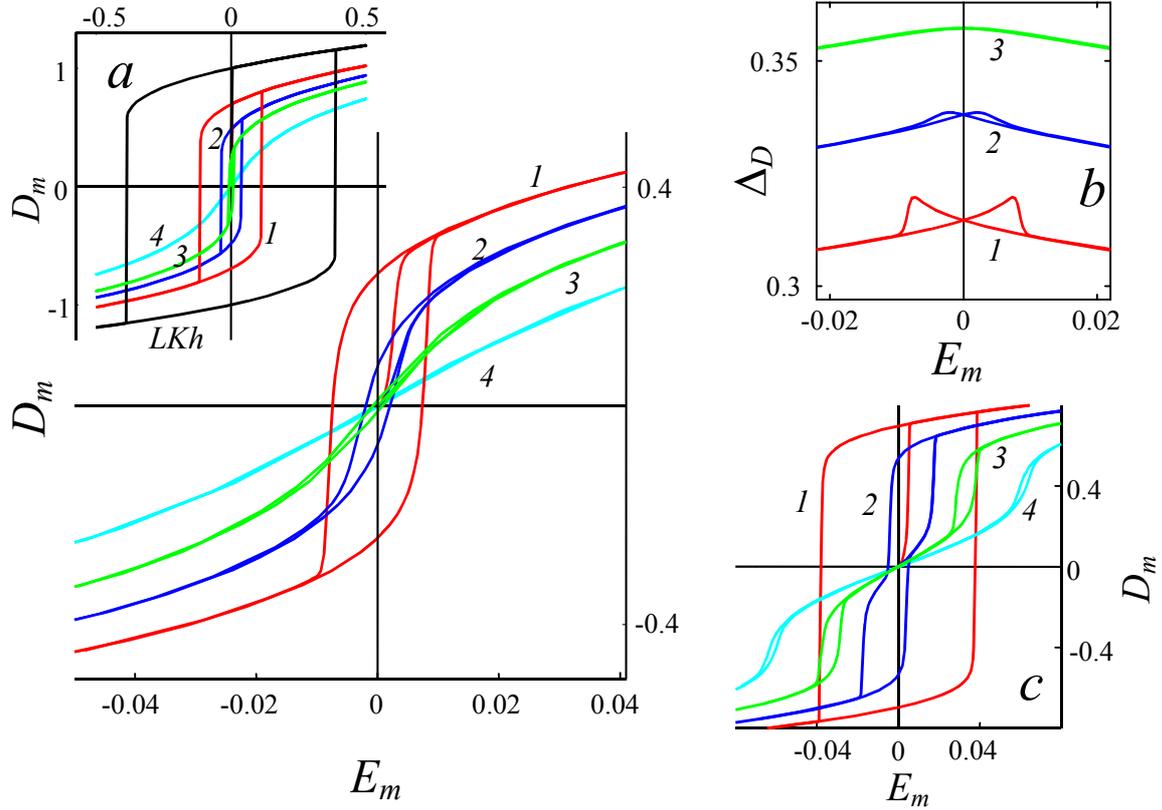

**Figure 3**. Hysteresis loops for $w$=10$^{-4}$, $R^2 = 0.1$ and different values of $\xi$ and $g$. **Basic plot** $D_m(E_m)$ and **Inset** $b$ $\Delta_D(E_m)$: $\xi$= -10 $g$=33 (curve1), 35 (curve 2), 37 (curve 3), 40 (curve 4). **Inset** $a$: $\xi$=-10, $g$=20 (curve 1), 28 (curve 2), 33 (curve 3), 45 (curve 4) and Landau-Khalatnikov loop (curve LKh). **Inset** $c$: $\xi$= -1 and $g$=4.5 (curve 1), 5 (curve 2), 5.5 (curve 3), 6 (curve 4).

The typical hysteresis loops obtained at 1≤g<10, negative $\xi$ value and different $R$ values are shown in Fig. 4. At small values $|\xi| \leq 1$ (see plot a), the increase of g value firstly leads to the coercive field decrease then to the constriction appearance, subsequent transformation to the double loop and finally to the loop disappearance. At high negative values $\xi \ll -1$ (see plot b) the loop becomes much slimmer and slightly lower under fluctuations $R$ increasing (compare Landau-Khalatnikov loop with other ones). Plot b demonstrates the drastic decrease of coercive field for $g$=5 and $R^2$ values more than 0.6.

Fig. 5 demonstrates the typical hysteresis loops at high negative $\xi$ values. The drastic decrease of coercive field corresponds to $\xi \sim -50$ at $g$=35 and $R^2 = 0.5$. The further increase of $\xi$ value leads to the coercive field increase, moreover the loop approaches the Landau-Khalatnikov one at $\xi \to -\infty$.

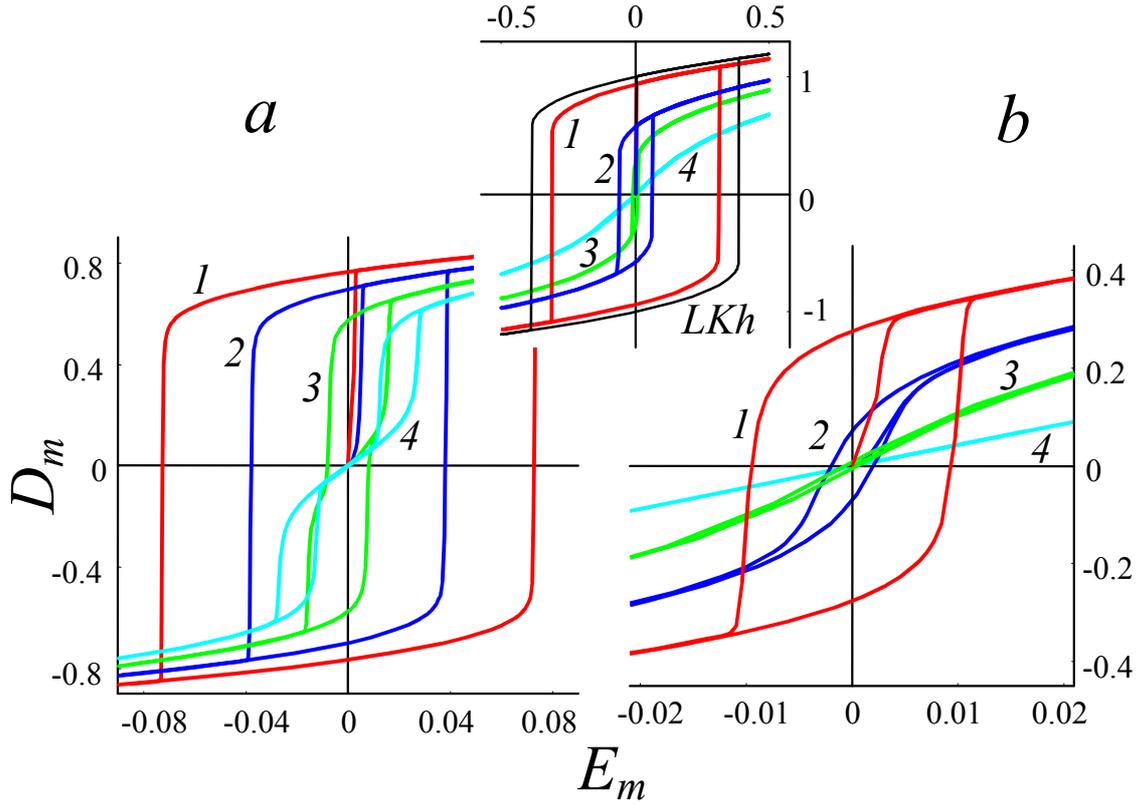

**Figure 4**. Hysteresis loops, for $w=10^{-4}$, $g=5$ and different values of $\xi$ and $R^2$. **Plot a**: $\xi$=-1, $R^2 = 0.08$ (curve 1), 0.09 (curve 2), 0.1 (curve 3), 0.105 (curve 4). **Plot b**: $\xi$=-10, $R^2 = 0.65$ (curve 1), 0.7 (curve 2), 0.75 (curve 3), 0.85 (curve 4). **Upper inset**: $\xi$=-10, $R^2 = 0.1$ (red curve 1), 0.5 (curve 2), 0.65 (curve 3), 1 (curve 4) and Landau-Khalatnikov loop ($R$=0, curve LKh).

One can notice from Figs.3-5 that the following scaling exists at $\xi << -1$: the calculated hysteresis loops reveal drastic decreasing of coercive field value at $-gR^2/\xi \geq 0.3$. In Appendix C we obtain the estimations for static remanent displacement $D_r \equiv \overline{D}(\omega = 0, E_0 = 0)$, linear dielectric permittivity $\varepsilon_r \equiv d\,\overline{D}(\omega = 0, E_0 = 0)/d\,E_0$ and coercive field value $E_C(\omega = 0)$, namely at $|\xi| >> 1$:

$$D_r \sim D_S\sqrt{1 + \frac{3gR^2}{\xi}}\,, \quad \varepsilon_r \sim \frac{1}{-2\alpha \cdot \left(1 + 3gR^2/\xi\right)}, \qquad |E_C| \sim \frac{2|\alpha D_S|}{3\sqrt{3}}\left(1 + \frac{3gR^2}{\xi}\right)^{3/2}. \quad (19)$$

Note, that (19) are valid for $R < 1$, $\xi < -10$, $g > 5$ with 10% accuracy. Really, at $\xi << -1$ $D_r$, $\varepsilon_r$ and $E_C$ depend on the combination $-gR^2/\xi \approx 16\pi^2\,d^2\overline{\delta\rho_s^2}/D_S^2$, with the transition point at $-gR^2/\xi \sim 0.3$.

We can conclude that the increasing of charged defect concentration (as well as its fluctuations) leads to the drastic decreasing of the coercive field, appearance of constricted and double hysteresis loops related to the inhomogeneous displacement switching. We demonstrate that this result much better agrees with experiments in thick films and bulk samples, than predictions of conventional Landau-Ginzburg and nucleation theories in the next section.

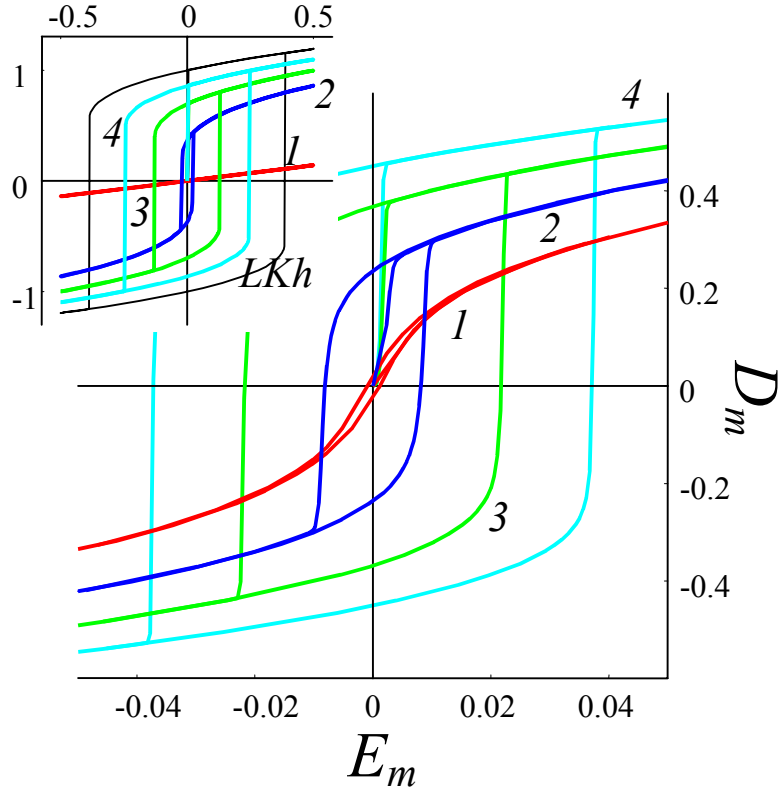

**Figure 5**. Hysteresis loops for $w=10^{-4}$, g=35, $R^2 = 0.5$ and different values of ξ: -50 (curve 1), -55 (curve 2), -60 (curve 3), -65 (curve 4). **Inset**: $\xi = -10$ (curve 1), -50 (curve 2), -100 (curve 3), -200 (curve 4) and Landau-Khalatnikov loop (curve LKh).

## 6. Comparison with experiment

Solid solutions Pb(Zr,Ti)O₃ (PZT) are widely used ferroelectrics, however much more work is needed to determine the physics of these materials [13]. A wide range of low level additives (0.0-5.0%) significantly influence on the dielectric properties of PZT at room temperatures. At liquid helium temperatures, all these "extrinsic" effects disappear and the experimental data agree with "single domain" intrinsic permittivity calculated from thermodynamic theory. Such donor additives as La, Nb, Nd or Ce at room temperature "soften" PZT dielectric properties, in particular hysteresis loops have rather high remanent displacement and low coercive field [13]. We suppose that aforementioned dopants as well as numerous unavoidable Pb vacancies (originated due to the high volatility of PbO, see chapter 10 in [12]) play a role of randomly distributed charged defects.

For thick Pb(Zr$_x$Ti$_{1-x}$)O₃ films ( $x \sim 0.5$, $\ell \geq 2 \mu m$ ) and bulk samples typical values are the following: remanent displacement $D_r = (20 \div 50) \mu C / cm^2$ and coercive field $E_C = (10 \div 100) kV / cm$ (see [13] and Tab.1). Usually the ferroelectric hysteresis loops of "soft" PZT films, obtained with the help of the conventional Sawyer-Tower circuit at low frequency ( $\omega \sim (0.1 \div 1) kHz$ ), are rather "thin" and "sloped" (see Figs. 6, 8).

For PZT Landau-Ginsburg free energy expansion coefficients $\alpha(x)$, $\beta(x)$ strongly depends over Zr molar fraction $x$. Near the morphotropic boundary $x \approx 0.5$ PZT is ferroelectric with the second order phase transition (β>0).

**Table 1**. Data for "soft" $Pb(Zr_xTi_{1-x})O_3$ films.

| Composition, Ref., method | Substrate, doping | Top electrode | Thickness, $\mu m$ | $\varepsilon_r (E=0)$ calc. from the loop | Ampl. $E_0$, kV/cm | Rem. displacement $D_r$, $\mu C/cm^2$ | Coercive field $E_C$, kV/cm |
|---|---|---|---|---|---|---|---|
| x=0.52, PZT-LQ [19] $\omega \sim 1kHz$ T=25°C ADM | SrRuO$_3$ / Si, Nb-modified | Ag, Au | 1.4 | 2800 | 400 | 28 | +96, -70 |
| | | | 20 | 3500 | 400 | ±28 | ±52 |
| | | | **Our fitting** (Fig. 6) | 2700 | 400 | ±25 | ±52 |
| x=0.54, [20] $\omega \sim 1kHz$ T=25°C RFMS | Pt/Si 1% Nb | Pt | 1.9 | ~3000 | 105 | +25 -28 | -23 +33.5 |
| | | | **Our fitting** (Fig. 8) | ~4000 | 105 | +20 -25 | -22.5 +32.5 |

For Landau-Ginsburg free energy $\alpha D^2/2 + \beta D^4/4 + \chi D^6/6 + ...$ expansion the following data are known: $\alpha(0.50 \div 0.54) = -(9.8 \div 12.2) \cdot 10^7 m/F$, $\beta(0.50 \div 0.54) = (19.1 \div 33.2) \cdot 10^7 m^5/C^2F$, (obtained by linear interpolation of data from Tab.7.1. in [13] at T=25°C, SI units). From these data it is easy to calculate Landau-Ginsburg remanent displacement $D_r = \sqrt{-\alpha/\beta} \approx (72 \div 61)\mu C/cm^2$ and static ($\omega = 0$) thermodynamic coercive field $E_C = 2\sqrt{-\alpha^3/27\beta} \approx (269 \div 284)kV/cm$ in $D^4$-approximation. These values do not match experimental data for $D_r$ and $E_C$ (see Tab.1). In order to improve calculations for $D_r$ higher coefficient $\chi(0.5 \div 0.54) = (8.0 \div 11.3) \cdot 10^8 m^7/C^4F$ is used [13], but even such approach leads to the significantly overestimated values of $E_C \geq 250\,kV/cm$.

The observed values of $D_r$ and $E_C$ depends not only over composition $x$, but over film thickness, fabrication method, substrate material, etc. The thickness-driven correlation effects [14] should be neglected if only the film thickness $\ell >> \gamma/\lambda|\alpha|$. For the typical values of surface extrapolation length $\lambda \leq 10^{-8}m$, gradient term $\gamma \leq 10^{-7}F \cdot m$ and $|\alpha| \sim 10^{+8}m/F$ far from Curie temperature [1], one obtains that thickness-driven effects become noticeable only for $\ell \leq 0.5\mu m$. So, for $\mu$m-thick PZT films we can neglect thickness-driven effects and use the results (17)-(19). However, up to 5$\mu$m films such interface phenomena as film-substrate mismatch and charged layers cause loops asymmetry [15] due to the induced electric field, and mismatch-induced homogeneous elastic stresses renormalize the free energy coefficients $\alpha$, $\beta$ [37].

In accordance with (18)-(19) we regard values $D_S = \sqrt{-\alpha/\beta}$, $E_S = 2\alpha D_S/3\sqrt{3}$ as hypothetical ones for perfect PZT at morphotropic boundary $x$=0.5, values $w << 1$ and $gR^2/\xi$ ($\xi << -1$) as fitting parameters determined by unavoidable cations vacancies depending over molar fraction $x$ and/or Nb$^{5+}$ concentration. Our fitting is presented in Figs.6-8. The discrepancy at high external field between calculated and measured the loops is regarded to the $D^4$-approximation used in our model. In order to improve the fitting the higher terms $D^6$, ... should be taken into account.

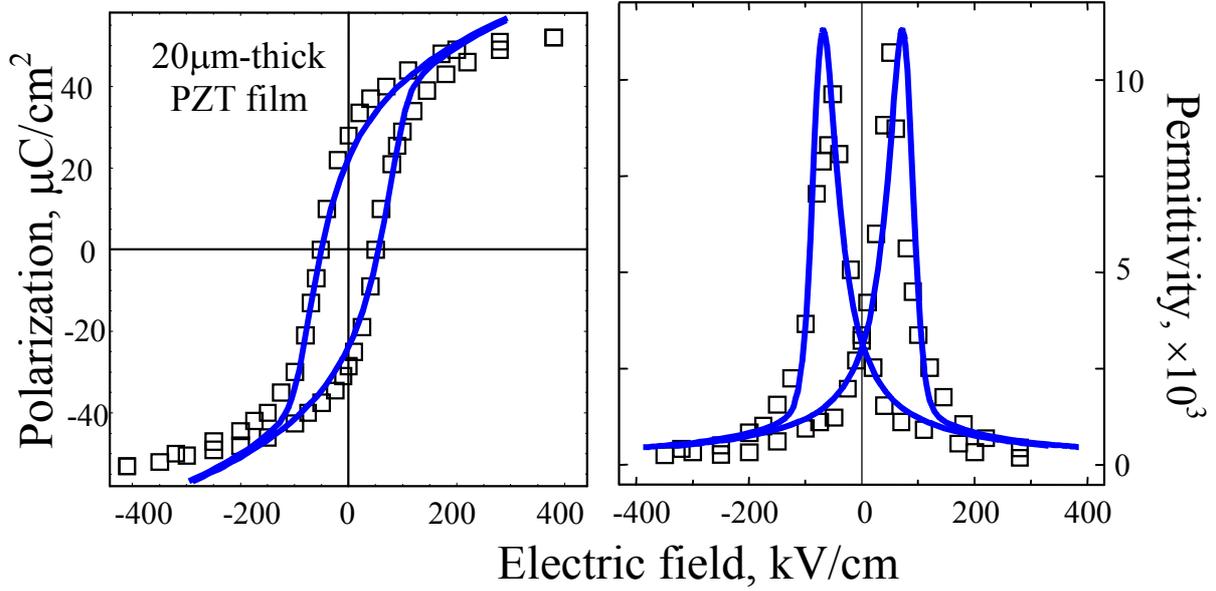

**Figure 6**. Hysteresis loop observed in a 20μm-thick Nb-modified PZT film fabricated by aerosol deposition method (ADM). Squares are experimental data from [19] at $E_0$=400k$V/cm$, solid curve is our fitting at $w = 0.017$, $\xi \leq -10$, $gR^2/\xi = -0.34$ (scaling region) and $D_S = 72\mu C/cm^2$, $E_S = 269\,kV/cm$ calculated from [13] at $x$=0.5.

The nucleation theory gives coercive field value independent over external field amplitude $E_0$. At low frequencies ($w \leq 0.1$) Landau-Khalatnikov equation gives coercive field values very close to thermodynamic limit $2\sqrt{-\alpha^3/27\beta}$ almost independent over external field amplitude $E_0$ (so called saturated "square" loops), whereas experiment data for PZT show its strong dependence over $E_0$ (see Fig.7b). In contrast to Landau-Khalatnikov model, our model shows quantitatively agreement with experiment (see Fig.7).

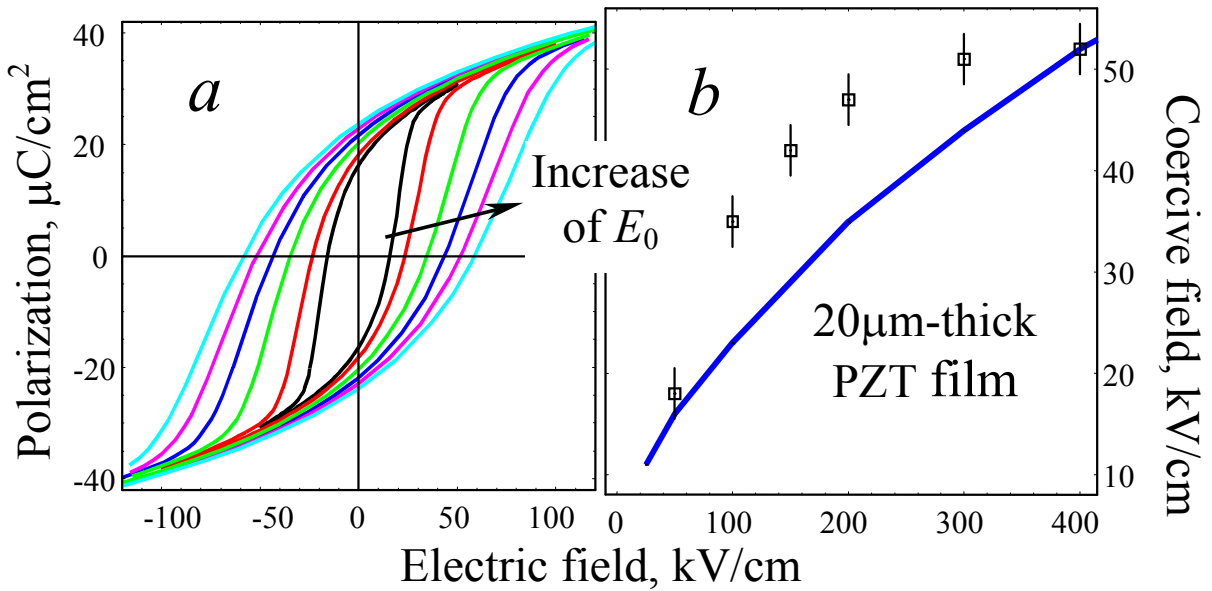

**Figure 7**. Hysteresis loop shape at different electric field amplitudes $E_0$: 50, 100,200, 300, 400 and 500kV/cm (plot **a**). Coercive field as a function of applied electric field amplitude (plot **b**). Squares are experimental data for 20μm-thick Nb-modified PZT film [19], solid line - our theory. Parameters are the same as in Fig.6.

The right-hand shift of the 1.9μm-thick PZT film is related to the laking layer near the boundary metal-semiconductor at the bottom electrode (see chapter 14 in [38] and Fig.8).

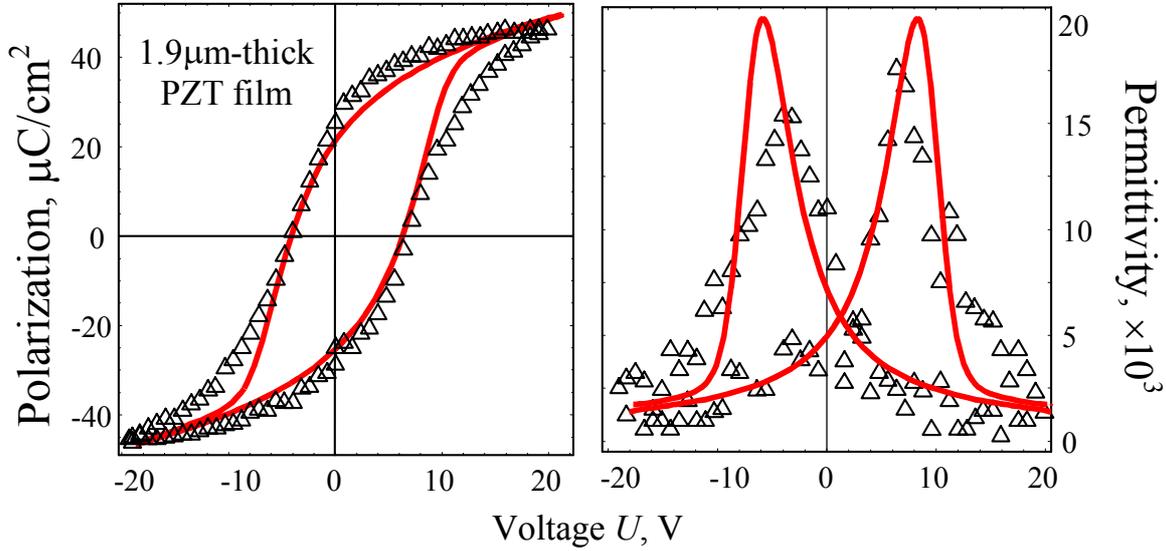

**Figure 8**. Centered hysteresis loop observed in a 1.9μm-thick "soft" PZT film manufactured by RF magnetron sputtering (RFMS). Triangles are experimental data given by authors of [20], solid curve is our fitting at $w = 0.02$, $gR^2/\xi = -0.36$ ($\xi = -4$, $R = 1$, $g = 1.45$) and $D_S = 72 \, \mu C / cm^2$, $E_S = 269 \, kV / cm$ ($U_S = 51.2 \, V$).

It is follows from Tab. 1 and Figs.6-8 that our model can quantitatively describe typical ferroelectric hysteresis loops in thick "soft" PZT films, in particular it gives correct coercive field values and its dependence over external field amplitude.

The constricted and double loops were observed in PZT ceramics doped with Nd [18], La [24], [25]. Sometimes constricted loops disappear after baking at high pressures, annealing in special atmospheres or several hundreds of switching cycles [18], [24], [25]. In order to explain this effect, we assume that ceramics treatment or fatigue can cause the significant changing of charged defects spatial distribution (e.g. the characteristic distance $d$ and concentration $n_d$). The latter leads to the changing of $gR^2$ and $\xi$ values in accordance with (18). Thus either appearance of constricted and double hysteresis loops in aged materials [18], [25] or their disappearance depends on the values of parameters $R$, $\xi$ and $g$ after sample treatment.

It should be noted, that the origin of the constricted or double loops in aged ferroelectric BaTiO$_3$ and (Pb,Ca)TiO$_3$ ceramics caused by the mechanical clamping of spontaneous polarization switching [12] and so it lies outside our theoretical consideration. This may be related to the fact that possible evolution of the charge fluctuations caused by the relaxation/origin of internal stresses around defects was not taken into account in our model.

Let us briefly compare our theoretical results with classical ones concerning lightly donor doped BaTiO$_3$. In many cases nonisovalent additives or unavoidable impurities containing even in very restricted quantity significantly attenuate piezoelectric properties of BaTiO$_3$ ceramics, increase electric conductivity, diffuse the peak of dielectric response and make hysteresis loop slim with low coercive field (see chapter 5 in [12]). That is why we consider the influence of charged defects on the ferroelectric properties of such "soft" materials, neglecting the contribution of inhomogeneous mechanical deformations arising due to intergranular stresses as well as local symmetry distortion appeared near the defects.

The (0.1-0.3)% La, Nb or Ce-doping changes wide band-gap intrinsic semiconductor BaTiO$_3$ to an extrinsic $n$-type semiconductor, achieved by replacing Ba$^{2+}$ or Ti$^{4+}$ with ions Le$^{3+}$, Ce$^{3+}$ or Nb$^{5+}$ with higher valency [12], [39]. In the most of BaTiO$_3$ ceramics such unavoidable impurity as Fe$^{3+}$ is present. In such case, the sample has slightly brown hue. Such ceramics reveals "soft" hysteresis loops with small area and low coercive field, diffuse dielectric properties and disappearance of spontaneous displacement (see [12], Tab. 2), in contrast to BaTiO$_3$ single crystals which undergoes first order phase transition with abrupt displacement disappearance at $T$=110°C. We supposed, that coupled equations (18) evolved for ferroelectrics with second order phase transition, can be applied to the doped or "soft" BaTiO$_3$ ceramics. Obtained results are summarized in Fig. 9 and Tab.2.

**Table 2**. Data for BaTiO$_3$ ceramics.

| Compo-sition, doping of bulk BaTiO$_3$ cera-mics | Experimental values Figs.5.16, 5.33 in [12], $T$=25°C, $\omega = 60 Hz$, $E_0 = 15 kV/cm$ | | Landau-Ginsburg theory for pure bulk BaTiO$_3$, **T=25°C** $\ell \to \infty$ | | Nucleation theory [6] $E_C \sim \left(\dfrac{1}{\ell}\right)^{2/9}$ | **Our fitting** at $\ell \to \infty$ (see Fig.9) | |
|---|---|---|---|---|---|---|---|
| | $E_C$, kV/cm | $D_r$, μC/cm$^2$ | $E_C$, kV/cm | $D_r$, μC/cm$^2$ | $E_C$, kV/cm | $E_C$, kV/cm | $D_r$, μC/cm$^2$ |
| "soft" | 3.5 | 7.5 | 200 | 25 | 1 ($\ell = 25\mu m$) | 4 | 7.5 |
| 3% Nb | 0.94 | 0.8 | | | 0 ($\ell \to \infty$) | 1 | 0.8 |

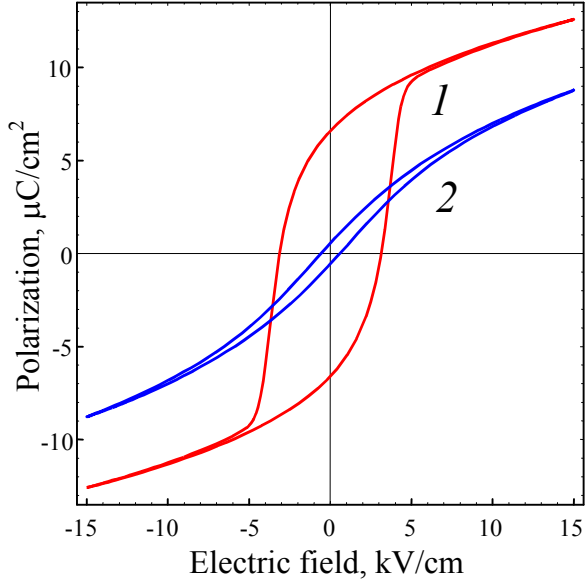

**Figure 9**. Our fitting for hysteresis loop observed in BaTiO$_3$ ceramics (Figs.5.16, 5.33 in [12]) obtained at $E_0$=15$kV/cm$, $\xi \leq -10$, $w = 0.005$, $E_S = 200 kV/cm$, $D_S = 25 \mu C/cm^2$, curve 1 -"soft" ceramics ($gR^2/\xi = -0.33$), curve 2 - ceramics doped with 3% of Nb ($gR^2/\xi = -0.39$).

It is follows from Tab. 2 and Fig.9 that our model can quantitatively describe typical ferroelectric hysteresis loops in "soft" BaTiO$_3$ ceramics, in particular it gives correct coercive field values.

### Conclusion

We have proposed the phenomenological description of polarization switching peculiarities in ferroelectric semiconductors with charged defects and prevailing extrinsic conductivity. Exactly we have modified Landau-Ginsburg approach for the aforementioned inhomogeneous systems and firstly obtained the system of coupled equations (17). Solving the system (17) one can get the information about system ordering as a whole, without defining space distribution of the appeared displacement inhomogeneities, however the present model has the following advances over the conventional ones and our previously published works [15], [34]-[35]:

• The coupled system (17) reveals **inhomogeneous** displacement switching in contrast to the Landau-Khalatnikov equation, which describes homogeneous one.

• We have shown that the increasing of charged defect concentration (as well as its fluctuations) leads to the drastic decreasing of the coercive field, appearance of constricted and double hysteresis loops.

• Obtained results quantitatively describe typical ferroelectric hysteresis loops in thick donor doped PZT films and BaTiO$_3$ ceramics, in particular our model gives correct coercive field values and its dependence over external field amplitude, in contrast to the conventional Landau-Ginsburg and nucleation theories.

### Acknowledgements

The author is greatly indebted to Prof. N.V. Morozovsky for frutfull discussions of the model and useful remarks to the manuscript and Prof. D.Remiens for avaliable experimental data.

### Appendix A

Let us estimate the dependence of parameters $\eta$ and $R^2$ on the charged defects distribution characteristics. Values of the average distance $d$ between them and size $r_0$ can be expressed via defect concentration $n_d$ and lattice constant $a$ as $d = a/\sqrt[3]{n_d}$ , $r_0 \sim a$ . If charged defects are clusterized distance $d$ can be several times greater than $1/\sqrt[3]{n_d}$ . We consider the case when defects concentration $n_d$ does not exceed several percents, i.e. $d > 3r_0$ (see Fig. 1). For the sake of simplicity, we approximate the defects charge densities by isotropic Gaussian distributions.

$$\rho_s(\mathbf{r}) = \sum_{i=1}^{N} \frac{q_i}{\pi^{3/2} r_{0i}^3} \exp\left[ -\frac{(\mathbf{r} - \mathbf{r}_i)^2}{r_{0i}^2} \right]. \tag{A.1a}$$

The charge density of carriers screening clouds originated around charge defects can be estimated in Gaussian approximation using the Debye potential [40] and Green function method, namely at $r_0 \sim R_D$ we obtain:

$$n(\mathbf{r}) \approx n_f + \sum_{i=1}^{N} \frac{e_i}{\pi^{3/2} (r_{0i} + R_D)^3} \exp\left[ -\frac{(\mathbf{r} - \mathbf{r}_i)^2}{(r_{0i} + R_D)^2} \right]. \tag{A.1b}$$

Here $n_f$ is the charge density of free electrons in the conduction band of extrinsic semiconductor [28], [38]. In our case high defects concentration $n_d \sim 1\%$ provides $n_f \approx \bar{n} = -\bar{\rho}_s$, and so only the relatively small amount of electron charge $e_i$ are localized near charged defects $q_i$.

In the hypothetical case, when all impurity atoms are identical and regularly distributed one obtains that $q_i \equiv q_0$, $e_i \equiv e_0$, $r_{0i} \equiv r_0$, $|\mathbf{r}_{i+1} - \mathbf{r}_i| \equiv d$ and thus we obtain from (A.1) that

$$\bar{\rho}_s = \frac{N}{V} \int_V d^3\mathbf{r} \frac{q_0}{\pi^{3/2} r_0^3} \exp\left( -\frac{r^2}{r_0^2} \right) = q_0 \frac{N}{V} \equiv \frac{q_0}{d^3}, \tag{A.2}$$

$$\overline{n} = n_f + \frac{N}{V}\int_V d^3\mathbf{r}\, \frac{e_0}{\pi^{3/2}(r_0+R_D)^3}\exp\left(-\frac{r^2}{(r_0+R_D)^2}\right) = n_f + \frac{e_0}{d^3}, \quad \text{(A.3)}$$

It is obvious from (A.2) and (A.3) that $q_0 = \overline{\rho}_s d^3$ and $e_0 = -\left(\overline{\rho}_s + n_f\right)d^3$ allowing for $-\overline{\rho}_s = \overline{n}$. Similarly to (A.2,3) we calculate that

$$\overline{\rho_s(\mathbf{r})n(\mathbf{r})} = \overline{\rho}_s n_f - \frac{\overline{\rho}_s\left(n_f+\overline{\rho}_s\right)}{\left[1+r_0^2/(r_0+R_D)^2\right]^{3/2}}\left(\frac{d}{\sqrt{\pi}r_0}\right)^3, \qquad \overline{\rho_s^2(\mathbf{r})} = \overline{\rho}_s^2\left(\frac{d}{\sqrt{2\pi}r_0}\right)^3. \quad \text{(A.4)}$$

Now it is easy to find the correlations

$$\overline{\delta\rho_s\,\delta n} \equiv \overline{\left(\rho_s-\overline{\rho}_s\right)\left(n-\overline{n}\right)} = \overline{\rho_s\,n} + \overline{\rho}_s^2, \qquad \overline{\delta\rho_s^2} \equiv \overline{\left(\rho_s-\overline{\rho}_s\right)^2} = \overline{\rho_s^2} - \overline{\rho}_s^2. \quad \text{(A.5)}$$

Really, with the help of above-mentioned assumptions and calculations (A.2)-(A.5) it is easy to obtain that

$$\overline{\delta\rho_s\,\delta n} = \overline{\rho}_s\left(n_f+\overline{\rho}_s\right)\cdot\left(\frac{\left(d/\sqrt{\pi}r_0\right)^3}{\left[1+r_0^2/(r_0+R_D)^2\right]^{3/2}}-1\right), \quad \overline{\delta\rho_s^2} = \overline{\rho}_s^2\left(\left(\frac{d}{\sqrt{2\pi}r_0}\right)^3-1\right). \quad \text{(A.6)}$$

Using the definitions $R^2 = -\overline{\delta\rho_s\delta n}/\overline{\rho}_s^2 \equiv \eta\,\overline{\delta\rho_s^2}/\overline{\rho}_s^2$ and $\eta \approx -\overline{\delta\rho_s\delta n}/\overline{\delta\rho_s^2}$, one derives from (A.6):

$$R^2 = \left(1+\frac{n_f}{\overline{\rho}_s}\right)\cdot\left(\frac{\left(d/\sqrt{\pi}r_0\right)^3}{\left[1+r_0^2/(r_0+R_D)^2\right]^{3/2}}-1\right), \qquad \eta \leq \left(1+\frac{n_f}{\overline{\rho}_s}\right)\left(\frac{2}{(1+R_D/r_0)^2+1}\right)^{3/2}. \quad \text{(A.7)}$$

Allowing for $n_f \approx -\overline{\rho}_s$, we obtain that always $\eta \ll 1$. Thus at $d \sim (5\div10)r_0$ and $\eta \leq 0.01$ $R^2 = \eta(7\div60) \leq 1$. However, when defects concentration becomes very small, e.g. $d\to\infty$ and $n_d \to 0$, intrinsic conductivity could not be neglected in comparison with extrinsic one and therefore the obtained expressions (A.6)-(A.7) become incorrect.

### Appendix B

The equations for $\overline{\delta D^2}$ and $\overline{\delta D\,\delta\rho_s}$ obtained directly from (12) have the form:

$$\frac{\Gamma}{2}\frac{\partial}{\partial t}\overline{\delta D^2} + \left(\alpha+3\beta\overline{D}^2(t)\right)\overline{\delta D^2} + \beta\,\overline{\delta D^4} = \gamma\,\overline{\delta D\frac{\partial^2\delta D}{\partial\mathbf{r}^2}} + \overline{\delta D\delta E_z}, \quad \text{(B.1)}$$

$$\Gamma\frac{\partial}{\partial t}\overline{\delta D\delta\rho_s} + \left(\alpha+3\beta\overline{D}^2(t)\right)\overline{\delta D\delta\rho_s} + \beta\,\overline{\delta D^3\,\delta\rho_s} = \gamma\,\overline{\delta\rho_s\frac{\partial^2\delta D}{\partial\mathbf{r}^2}} + \overline{\delta\rho_s\,\delta E_z}. \quad \text{(B.2)}$$

Under multiplying (12) over $\delta D^2$ or $\delta\rho_s\delta D$ and averaging allowing for (5) - (6) one obtains that

$$\overline{\delta D^4} = \left(\overline{\delta D^2}\right)^2, \qquad \overline{\delta D^3\,\delta\rho_s} = \overline{\delta D^2}\;\overline{\delta D\delta\rho_s}. \quad \text{(B.3)}$$

Taking into account that the average period of the inhomogeneities distribution is $d$ (see Fig. 1), one obtains that $\partial/\partial\mathbf{r} \sim i/d$ and so

$$\gamma\,\overline{\delta D\frac{\partial^2\delta D}{\partial\mathbf{r}^2}} \sim -\frac{\gamma}{d^2}\overline{\delta D^2}, \qquad \gamma\,\overline{\delta\rho_s\frac{\partial^2\delta D}{\partial\mathbf{r}^2}} \sim -\frac{\gamma}{d^2}\overline{\delta\rho_s\delta D}. \quad \text{(B.4)}$$

Let us express the field variation $\delta E_z$ via $\delta D$ and $\delta\rho_s$. In accordance with (17):

$$\delta E_z \approx \frac{\partial}{\partial t} \frac{\delta D}{4\pi\mu\,\overline{\rho}_s} - E_0(t)\left(\frac{\delta\rho_s}{\overline{\rho}_s} - \frac{div(\delta\mathbf{D})}{4\pi\overline{\rho}_s}\right) + \frac{\kappa}{\mu}\frac{\partial}{\partial z}\frac{\delta\rho_s}{\overline{\rho}_s} - \frac{\kappa}{\mu}\frac{\partial}{\partial z}\frac{div(\delta\mathbf{D})}{4\pi\overline{\rho}_s} + \frac{\delta n\,\delta E_z - \overline{\delta n\,\delta E_z}}{\overline{\rho}_s}$$

$$(B.5)$$

For a thick sample with equivalent boundaries $z = \pm\ell$ the term $\overline{\dfrac{\partial}{\partial z}\delta D^2}$ is equal to zero. In accordance with comments to (6) $\overline{\delta n\,\delta E_z^2} = 0$ and $\overline{\delta\rho_s\,\delta n\,\delta E_z} = 0$, thus one can derive from (B.5) the following approximations for the correlations:

$$\overline{\delta D\delta E_z} = \frac{1}{8\pi\mu\,\overline{\rho}_s}\frac{\partial}{\partial t}\overline{\delta D^2} - E_0(t)\frac{\overline{(\delta\rho_s\delta D)}}{\overline{\rho}_s} + \frac{\kappa}{\mu}\overline{\left(\delta D\frac{\partial}{\partial z}\frac{\delta\rho_s}{\overline{\rho}_s}\right)} - \frac{\kappa}{4\pi\overline{\rho}_s\mu}\overline{\delta D\frac{\partial}{\partial z}div(\delta\mathbf{D})}. \quad \text{(B.6)}$$

In accordance with (13) the term $\dfrac{\kappa}{\mu}\overline{\left(\delta D\dfrac{\partial}{\partial z}\dfrac{\delta\rho_s}{\overline{\rho}_s}\right)} = -\dfrac{\kappa}{\mu\overline{\rho}_s}\overline{\left(\delta\rho_s\dfrac{\partial}{\partial z}\delta D\right)}$ can be estimated as $-\dfrac{4\pi\,\kappa}{\mu\overline{\rho}_s}\overline{\delta\rho_s(\delta\rho_s + \delta n)} \approx -(1-\eta)\dfrac{4\pi\,\kappa}{\mu\overline{\rho}_s}\overline{\delta\rho_s^2}$. In accordance with the definition of Debye radius $R_D = \sqrt{\kappa/4\pi n\mu}$, this term $-(1-\eta)\dfrac{4\pi\,\kappa}{\mu\overline{\rho}_s}\overline{\delta\rho_s^2} = 16\pi^2 R_D^2(1-\eta)\overline{\delta\rho_s^2}$. Taking into account that the average period of the inhomogeneities distribution is $d$ (see Fig. 1), the last term in the right-hand side of (B.6) can be estimated as $\dfrac{\kappa}{4\pi\overline{\rho}_s\mu}\overline{\dfrac{\partial\delta D}{\partial z}div(\delta\mathbf{D})} \sim \dfrac{\kappa}{4\pi\overline{\rho}_s\mu}\overline{\left(\dfrac{\partial\delta D}{\partial z}\right)^2} = -\dfrac{R_D^2}{d^2}\overline{\delta D^2}$. Thus the correlation

$$\overline{\delta D\delta E_z} = \frac{1}{8\pi\mu\,\overline{\rho}_s}\frac{\partial}{\partial t}\overline{\delta D^2} - E_0(t)\frac{\overline{(\delta\rho_s\delta D)}}{\overline{\rho}_s} + 16\pi^2 R_D^2(1-\eta)\overline{\delta\rho_s^2} - \frac{R_D^2}{d^2}\overline{\delta D^2}. \quad \text{(B.7)}$$

For a thick sample with equivalent boundaries $z = \pm\ell$ the term $\overline{\left(\dfrac{\partial}{\partial z}\dfrac{\delta\rho_s^2}{\overline{\rho}_s}\right)}$ is equal to zero and therefore we obtain from (B.5) that

$$\overline{\delta\rho_s\delta E_z} = \frac{1}{4\pi\mu\,\overline{\rho}_s}\frac{\partial}{\partial t}\overline{\delta\rho_s\delta D} + E_0(t)\frac{\overline{\delta\rho_s\delta n}}{\overline{\rho}_s} - \frac{\kappa}{4\pi\overline{\rho}_s\mu}\overline{\delta\rho_s\frac{\partial}{\partial z}div(\delta\mathbf{D})}. \quad \text{(B.8)}$$

In accordance with (13) the last two terms in the right-hand side of (B.6) can be estimated as $E_0(t)\dfrac{\overline{\delta\rho_s\delta n}}{\overline{\rho}_s} \approx -\eta\dfrac{\overline{\delta\rho_s^2}}{\overline{\rho}_s}E_0(t)$, $\dfrac{\kappa}{4\pi\overline{\rho}_s\mu}\overline{\delta\rho_s\dfrac{\partial}{\partial z}div(\delta\mathbf{D})} \sim -\dfrac{\kappa}{4\pi\overline{\rho}_s\mu d^2}\overline{\delta\rho_s\delta D} = \dfrac{R_D^2}{d^2}\overline{\delta\rho_s\delta D}$ and so:

$$\overline{\delta\rho_s\delta E_z} = \frac{1}{4\pi\mu\,\overline{\rho}_s}\frac{\partial}{\partial t}\overline{\delta\rho_s\delta D} - \eta\frac{\overline{\delta\rho_s^2}}{\overline{\rho}_s}E_0(t) - \frac{R_D^2}{d^2}\overline{\delta\rho_s\delta D}. \quad \text{(B.9)}$$

Thus gradient terms in (B.1)-(B.2) can be either neglected at $(\gamma + R_D^2)/d^2 << \alpha$ or the coefficient $\alpha$ can be renormalized as $\alpha \to \alpha_R = (\alpha + (\gamma + R_D^2)/d^2)$. Using (B.3), (B.4), (B.7), (B.9) we obtain the equations (18b) and (18c) from the equations (B.1) and (B.2).

### Appendix C

In the stationary case ($\omega \to 0$, $\partial/\partial t \equiv 0$) the correlation $\overline{\delta D\delta\rho_s}$ can be expressed via $\overline{D^2}$ and $\overline{\delta D^2}$ from (17c), thus the equation (17b) acquires the form:

$$\overline{\delta D^2} = \frac{E_0^2 R^2}{\left(\alpha_R + 3\beta\overline{D}^2 + \beta\overline{\delta D^2}\right)^2} + \frac{\vartheta}{\left(\alpha_R + 3\beta\overline{D}^2 + \beta\overline{\delta D^2}\right)},$$ (C-1)

As it follows from (17a) $\overline{\delta D^2} = \left(E_0 - \alpha\overline{D} - \beta\overline{D}^3\right)\big/3\beta\overline{D}$, and the modified equation for $\overline{D}$ can be obtained directly from (C-1), namely

$$\left[\alpha + \frac{3\beta\left(E_0^2 R^2 + \vartheta\left(\alpha_R - \alpha/3 + 8\beta\overline{D}^2/3 + E_0/3\overline{D}\right)\right)}{\left(\alpha_R - \alpha/3 + 8\beta\overline{D}^2/3 + E_0/3\overline{D}\right)^2}\right]\overline{D} + \beta\overline{D}^3 = E_0.$$ (C-2)

Let us remind, that $\vartheta = \left(4\pi R_D\right)^2\left(1 - \eta\right)\overline{\delta\rho_s^2}$, $D_S^2 = -\alpha/\beta$, $R^2 = \eta\,\overline{\delta\rho_s^2}\big/\overline{\rho_s^2}$ (see (18)).

1) The remanent (spontaneous) displacement $D_r \equiv \overline{D}(E_0 = 0)$ obtained from (C-2) satisfies the following biquadratic equation

$$D_r^2 = D_S^2\left[1 - \frac{3\vartheta/D_S^2}{\left(\alpha_R - \alpha/3 + 8\beta D_r^2/3\right)}\right].$$ (C-3)

As it should be expected, the value $D_r^2 = D_S^2$ obtained in the Landau-Ginzburg model, is the zero approximation in (C-3) over parameter $\vartheta\big/\left(\alpha D_S^2\right)$. Thus, in the first approximation over parameter $\vartheta\big/\left(\alpha D_S^2\right)$ one derives from (C-3) that

$$D_r^2 \approx D_S^2\left[1 - \frac{3\,\vartheta/D_S^2}{\alpha_R - 3\alpha}\right].$$ (C-4)

2) The static linear dielectric permittivity $\varepsilon_r \equiv d\,\overline{D}(E_0 = 0)\big/d\,E_0$ obtained from (C-2) satisfies the following linear equation

$$\varepsilon_r\left[\alpha + \frac{3\beta\vartheta}{\left(\alpha_R - \alpha/3 + 8\beta D_r^2/3\right)} - \frac{16\beta^2\vartheta D_r^2}{\left(\alpha_R - \alpha/3 + 8\beta D_r^2/3\right)^2} + 3\beta D_r^2\right] = 1 + \frac{\beta\vartheta}{\left(\alpha_R - \alpha/3 + 8\beta D_r^2/3\right)^2}$$

In accordance with (C-3) $\alpha + \dfrac{3\beta\vartheta}{\left(\alpha_R - \alpha/3 + 8\beta D_r^2/3\right)} = -\beta D_r^2$, and thus we obtain

$$\varepsilon_r = \frac{1 + \dfrac{\beta\vartheta}{\left(\alpha_R - \alpha/3 + 8\beta D_r^2/3\right)^2}}{2\beta D_r^2 - \dfrac{16\beta^2\vartheta D_r^2}{\left(\alpha_R - \alpha/3 + 8\beta D_r^2/3\right)^2}}$$ (C-5)

As it should be expected, the value $\varepsilon_R \equiv 1\big/\left(-2\alpha\right)$ obtained in the Landau-Ginzburg model, is the zero approximation in (C-5) over parameter $\vartheta\big/\left(\alpha D_S^2\right)$. Thus, in the first approximation over parameter $\vartheta\big/\left(\alpha D_S^2\right)$ one derives from (C-4) and (C-5) that

$$\varepsilon_r \approx \frac{1}{-2\alpha}\cdot\frac{1}{1 + \dfrac{3\beta\vartheta}{\left(\alpha_R - 3\alpha\right)\alpha} - \dfrac{9\beta\vartheta}{\left(\alpha_R - 3\alpha\right)^2}}$$ (C-6)

3) The static coercive field value $E_C(\overline{D} = 0)$ can be determined from the divergence of generalized permittivity, namely from the condition $d\,\overline{D}(E_0 = E_C)\big/d\,E_0 \to \infty$. Thus, the condition of zero denominator in the cumbersome expression for $d\,\overline{D}(E_0)\big/d\,E_0$ obtained directly from (C-2), gives the following equation for coercive field value determination:

$$E_C = -2\beta\overline{D}^3 + \frac{\beta\left(16\beta\,\overline{D}^3 - E_C\right)\left(2E_C^2 R^2 + 9\left(\alpha_R - \alpha/3 + 8\beta\overline{D}^2/3 + E_C/3\overline{D}\right)\right)}{\left(\alpha_R - \alpha/3 + 8\beta\overline{D}^2/3 + E_C/3\overline{D}\right)^3}. \qquad \text{(C-7)}$$

The (C-7) is solved coupled with (C-2) at $E_0 = E_C$, and the couple of values $\left\{\overline{D}(E_C),\ E_C\right\}$ can be found at least numerically. As it should be expected, the values $\left\{\overline{D} = D_S/\sqrt{3},\quad E_C = -2\beta\overline{D}^3 \equiv \dfrac{2\alpha D_S}{3\sqrt{3}}\right\}$ obtained in the Landau-Ginzburg model, is the zero approximation in (C-7), (C-2) over parameters $R$, $9/\left(\alpha D_S^2\right)$. Thus, in the first approximation over these parameters one derives from (C-2) and (C-7) the following system

$$E_C \approx \alpha\overline{D}(E_C)\cdot\left[1 + \nu_1\right] + \beta\overline{D}^3(E_C), \quad E_C \approx -2\beta\overline{D}^3(E_C)\cdot\left[1 + \nu_2\right], \qquad \text{(C-8a)}$$

$$\nu_1 = -\frac{4\alpha^2 R^2}{9\left(\alpha_R - \alpha\right)^2} + \frac{3\beta 9}{\left(\alpha_R - \alpha\right)\alpha}, \quad \nu_2 = \frac{8\alpha^3 R^2}{3\left(\alpha_R - \alpha\right)^3} - \frac{9\beta 9}{\left(\alpha_R - \alpha\right)^2}. \qquad \text{(C-8b)}$$

In the first approximation over parameters $\nu_{1,2}$, one derives from the system (C-8) that $E_C = \dfrac{2\alpha D_S}{3\sqrt{3}}\left(1 + \dfrac{3\nu_1}{2}\right)$, $\overline{D}(E_C) = \dfrac{D_S}{\sqrt{3}}\left(1 + \dfrac{\nu_1}{2} - \dfrac{\nu_2}{3}\right)$. Allowing for (C-8b) we obtain:

$$E_C \approx \frac{2\alpha D_S}{3\sqrt{3}}\left(1 - \frac{2\alpha^2 R^2}{3\left(\alpha_R - \alpha\right)^2} + \frac{9\beta 9}{2\left(\alpha_R - \alpha\right)\alpha}\right). \qquad \text{(C-9)}$$

Let us rewrite the approximate formulas (C-4), (C-6) and (C-9) in the dimensionless variables $R$, $9/\left(\alpha D_S^2\right) = gR^2$ $\xi = \alpha_R/\alpha$ ($\alpha < 0$, $\alpha_R \geq 0$, see also (18)), namely:

$$D_r \approx D_S\sqrt{1 - \frac{gR^2}{1 - \xi/3}}, \qquad \varepsilon_r \approx \frac{-1/2\alpha}{1 - \dfrac{gR^2}{1 - \xi/3} - \dfrac{gR^2}{\left(1 - \xi/3\right)^2}}, \qquad \text{(C-10)}$$

$$E_C \approx \frac{2\alpha D_S}{3\sqrt{3}}\left(1 - \frac{2R^2}{3\left(1 - \xi\right)^2} - \frac{9gR^2}{2\left(1 - \xi\right)}\right). \qquad \text{(C-11)}$$

Comparing the approximate formulas (C-11), (C-12) with the numerical calculations based on (18) at $\omega \to 0$, i.e. on (C-3), (C-5) and (C-7), we obtain, that (C-10)-(C-12) are valid for $R < 1$, $\xi < -5$, $g > 5$ with 10% accuracy (e.g. see Fig 3). Moreover, the following estimations are valid

$$\left|E_C\right| \approx \frac{2\left|\alpha D_S\right|}{3\sqrt{3}}\left(1 - \frac{3gR^2}{\left(1 - \xi\right)}\right)^{3/2}, \qquad \varepsilon_r \approx \frac{1}{-2\alpha\cdot\left(1 - gR^2/\left(1 - \xi/3\right)\right)}. \qquad \text{(C.12)}$$